\documentclass[twoside]{article}
\usepackage{fleqn,espcrc2,epsfig,rotating}

\title{Nonperturbative tuning of $O(a^2)$ improved staggered fermions}

\author{Massimo Di Pierro\address[FNAL]{
Fermilab, PO Box 500, Batavia, IL 60563, USA}%
        \thanks{Poster presented at Lattice 2001, Berlin} and
	Paul Mackenzie\addressmark[FNAL]}       

\begin{document}

\begin{abstract}
We perform a nonperturbative tuning of the 
coefficients  in the $O(a^2)$ improved action for staggered
fermions. The mass splitting for the  pions of different doubler flavor
 is used as a measure 
of the symmetry breaking effects introduced by $O(a^2)$ discretization errors. 
We find that the flavor nondegeneracy can be somewhat reduced but  not
eliminated by such a tuning, indicating the need for new terms in the action
to reduce the nondegeneracy.
\vspace{1pc}
\end{abstract}

\maketitle

\section{INTRODUCTION}

Staggered fermions offer the possibility of doing unquenched calculations
on current computers with far less simulation time than Wilson type fermions.
In their simplest form, however,
they suffer from several well-known problems
which must be addressed before they can be used effectively~\cite{tro01}.
One significant problem with ordinary staggered fermions is the large
flavor nondegeneracy, worst in the pion sector. 
At tree level, it arises from transitions between
doubler quarks of different flavors induced by gluons of momentum 
$\pi$~\cite{lep98,leg98}.
In Ref.~\cite{lep99}, Lepage showed how to turn the ``fat-link'' 
improvement of the MILC collaboration~\cite{org98},
which suppresses the coupling of quarks to these gluons,
into a tree level ${\cal O}(a^2)$
improved action by proper tuning of the coefficients and the inclusion of 
an additional term.

Monte Carlo calculations
 showed a large reduction in pion flavor nondegeneracy with this
action~\cite{org99}. Significant  
flavor breaking still remains, however, so further improvement is desirable for
high precision calculations.
In this work, we perform a nonperturbative determination of gluonic 
corrections to the tadpole improved tree level $a^2$ improved staggered
action (called the ``Asqtad'' action by the MILC collaboration).
We find a small improvement in pion flavor breaking, but not the large 
reduction that is still desirable.  Therefore, incorporation of additional
operators into the action is needed~\cite{tro01}.

\section{THE IMPROVED STAGGERED ACTION}

We define a modified Asqtad 
 (mAsqtad) action, built in two steps: 
First, from the naive fermionic action, one modifies the definition of 
the covariant derivative:
\begin{eqnarray}
&&\!\!\!\!\!\!\!\!\!\!\!\!D^{\{c_i\}}_\mu \psi(x) = V_\mu \psi(x+\mu) - V_{-\mu}\psi(x-\mu) 
\label{fullyimp} \\
&&\!\!\!\!\!\!\!\!\!-\frac{1+c_5}{24u_0^2}\left[
(U_\mu)^3 \psi(x+3\mu)- (U_{-\mu})^3 \psi(x-3\mu)
\right]
\nonumber  
\end{eqnarray}
where $V_\mu$ is a fat link defined as
\begin{eqnarray}
&&\!\!\!\!\!\!\!\!\!\!\!\!V_\mu \equiv \frac{5}{8}(1+c_0) U_\mu \nonumber \\
&&\!\!\!\!\!\!\!\!\!+\frac{1+c_1}{16u_0^2} \sum_{\nu} 
U_{\pm\nu} U_\mu U_{\mp\nu}+\nonumber \\
&&\!\!\!\!\!\!\!\!\!+\frac{1+c_2}{64u_0^4} \sum_{\nu,\rho}
U_{\pm\rho} U_{\pm\nu} U_\mu U_{\mp\nu} U_{\mp\rho} \nonumber \\
&&\!\!\!\!\!\!\!\!\!+\frac{1+c_3}{384u_0^6} \sum_{\nu,\rho,\sigma}
U_{\pm\sigma} U_{\pm\rho} U_{\pm\nu} U_\mu 
U_{\mp\nu} U_{\mp\rho} U_{\mp\sigma} \nonumber \\
&&\!\!\!\!\!\!\!\!\!-\frac{1+c_4}{16u_0^4} \sum_{\nu} (U_{\pm\nu})^2 U_\mu (U_{\mp\nu})^2 
\end{eqnarray}
(the indices in the sums are always different from $\mu$ and among each other).
The choice of the coefficients $c_i = 0$ corresponds to the Asqtad action.

Second, the fermion $\phi(x)$ is mapped into a scalar field $\chi(x')$ by 
the relation
\begin{equation}
\phi(x)^a_\alpha=\sum_{A \in [2^4]} (\gamma_1^{A_1}\gamma_2^{A_2}\gamma_3^{A_3}\gamma_4^{A_4})^a_\alpha \chi(x+A)
\label{fermionicmap}
\end{equation}
where $\alpha$ is the spin index, $a$ is the flavor index, $[2^4]$ is a 4D hypercube. Note that eq.~(\ref{fermionicmap}) is invertible only if the fermion and the scalar live on different lattices, i.e. if the former lives on the blocked lattice of the latter~\cite{mesons}. We use this prescription to build the
${\bf 15}$-plet of pions of $SU(4)$ flavor.

Because of the remnant discrete flavor symmetry there are 
only seven inequivalent pions. 
We identify the seven inequivalent pions by their SU(4) 
flavor structure 
\begin{equation}
\xi=\gamma^5,\ \gamma^0 \gamma^5,\ \gamma^3 \gamma^5,\ \gamma^1 \gamma^2,\ \gamma^3 \gamma^4,\ \gamma^3, \gamma^3
\end{equation}
Our goal is that of tuning the coefficients $c_i$ around zero to reduce the 
mass splitting among these pions.

\begin{figure}
\begin{center}
\begin{turn}{270}
\epsfig{file=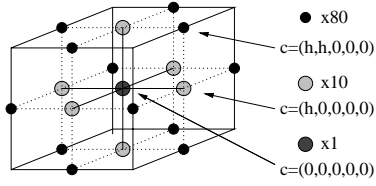, height=6cm}
\end{turn}
\vskip -1cm
\end{center}              
\caption{Three dimensional representation of the five dimensional space of coefficients and the points we considered.\label{5dcube}}
\end{figure}

\section{Computation}

Our computation was performed on 113 $O(a^2)$ improved gauge configurations at $\beta=7.4$ and $u_0=0.8629$ on a $24\times 8^3$ lattice. 
We preceded in the following way:

We chose a finite set of points in the space of the coefficients. For each 
point, we performed a lattice computation of the Goldstone pion and we fine tuned the quark mass $m$ in order to reproduce a mass for the Goldstone 
pion, $M_{\gamma^5}$, equal  to $(0.49 \pm 0.01)a^{-1}$ (this is an arbitrary number). This fine tuning is required since the mass renormalizes in different ways for the different choices of coefficients and we need to impose a physical renormalization condition. Then, for each point, we computed the whole spectrum of pions using the corresponding fine tuned value for the quark mass.

We chose 91 points in the space of coefficients, namely $c_i= 0$, $c_i=\pm h \delta_{ij}$, $c_i=\pm h \delta_{ij} \pm h \delta_{ik}$ for every value of $j$ and $k \neq j$ ($h=0.5$). These points are represented in Fig.~\ref{5dcube}. This choice enables us to evaluate numerically the first ($M'_{\xi,i}$) and second derivative ($M''_{\xi,ij}$) of the pion masses in respect to each coefficient of the action (keeping fixed the renormalization condition, i.e. the mass of the Goldstone pion). 

This work amounts to more than 10000 fermionic inversions of the action and more than 2000 fits of pion propagators; it was performed on the Fermilab QCD80 cluster. 
 
\section{Results}

\begin{figure}
\begin{center}
\epsfig{file=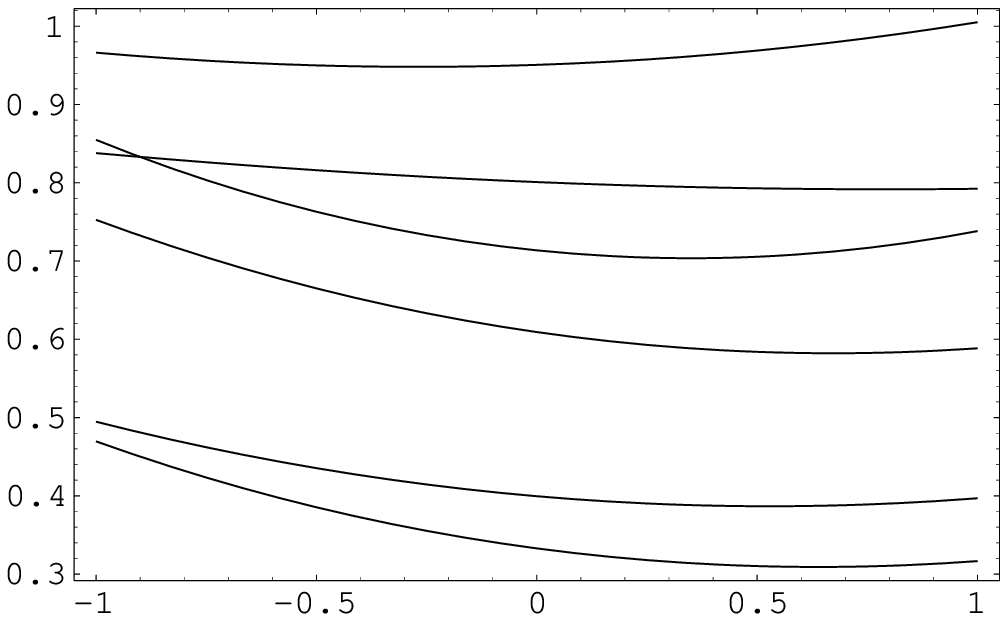, height=3.5cm}
$c_1$ \vskip 2mm
\epsfig{file=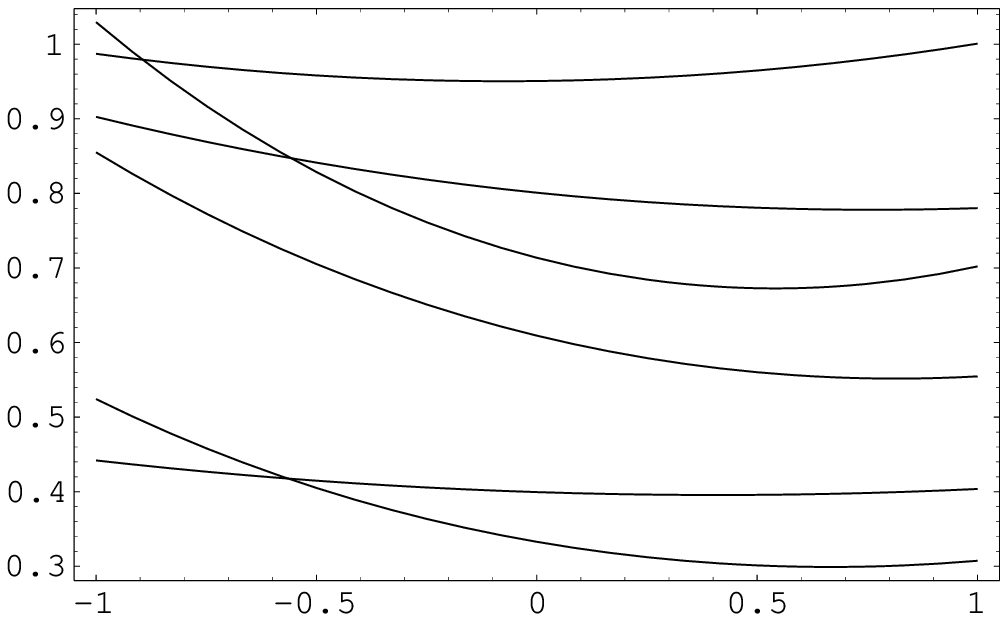, height=3.5cm}
$c_2$ \vskip 2mm
\epsfig{file=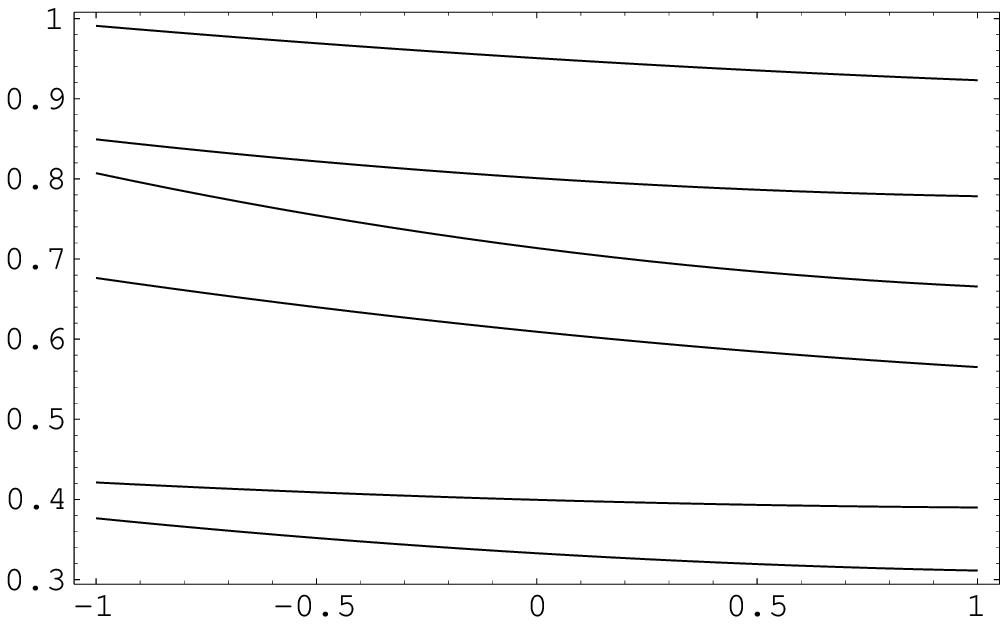, height=3.5cm}
$c_3$ \vskip 2mm
\epsfig{file=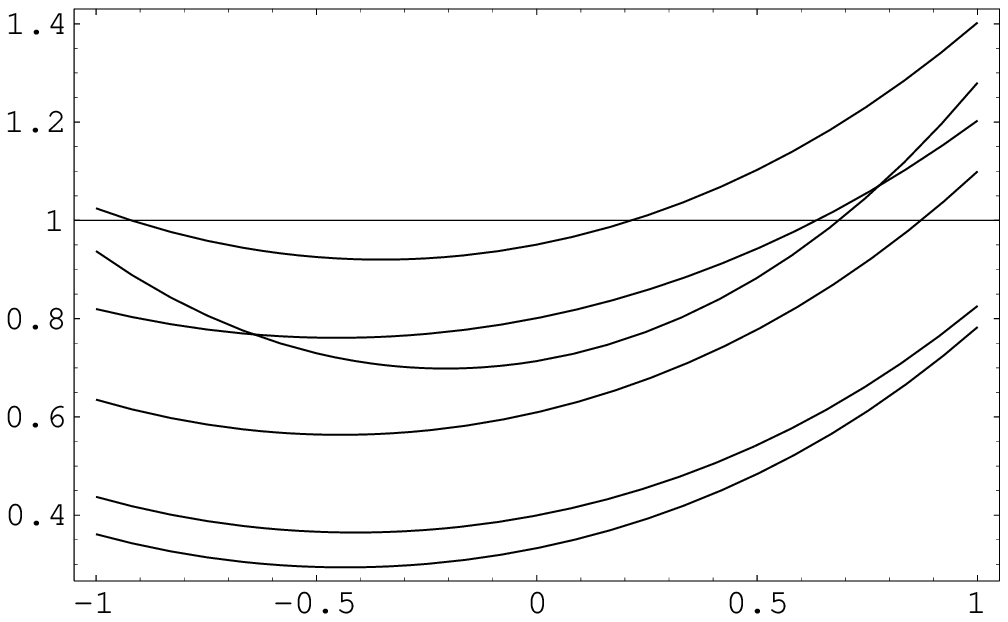, height=3.5cm}
$c_4$ \vskip 2mm
\epsfig{file=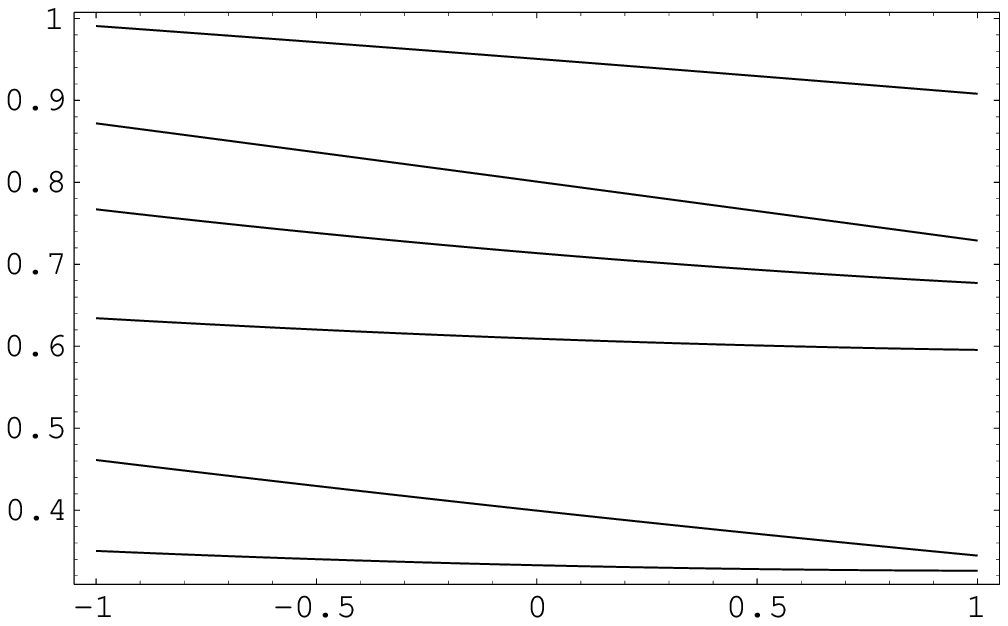, height=3.5cm}
$c_5$ 
\vskip -1cm
\end{center}
\caption{$F_\xi$ as function of $\{c_i\}$.\label{plotall}}
\end{figure}

We expand the pion masses as functions of the coefficients in the action, in Taylor series up to second order
\begin{equation}
M_\xi(c_i)=M^0_\xi+\sum_i M'_{\xi,i} c_i + \frac12 \sum_{i,j} M''_{\xi,ij}c_i c_j
\end{equation}
and we define
\begin{equation}
F_\xi(c_i)=M^2_\xi(c_i)-M^2_{\gamma^5}
\end{equation}
where $M_{\gamma^5}=const.$ because of our choice of renormalization condition.
In Fig.~\ref{plotall} we report sections of the function $F_\xi$. Each plot shows the value of $F$ of all pions, $\xi$, when we vary one single coefficient.

The first result that is already visible from the plots is that the 
spectrum is very mildly dependent on the action and there is no obvious 
direction in the space of coefficients where the pion mass splitting gets 
reduced for all pions at once. 

We report here the values for the meson masses corresponding to the 
Asqtad action (for a fine tuned light quark mass of $m =0.0347a^{-1}$)
\begin{eqnarray}
&\!\!\!\!\!M^0_{\gamma^5\phantom{\gamma^5}} = 0.592 \pm 0.003 \ \ 
&\!\!\!\!\!M^0_{\gamma^0\gamma^5} = 0.801 \pm 0.017 \nonumber \\
&\!\!\!\!\!M^0_{\gamma^3\gamma^5} = 0.758 \pm 0.006 \ \ 
&\!\!\!\!\!M^0_{\gamma^1\gamma^2} = 1.021 \pm 0.030 \nonumber \\
&\!\!\!\!\!M^0_{\gamma^3\gamma^4} = 0.923 \pm 0.012 \ \  
&\!\!\!\!\!M^0_{\gamma^3\phantom{\gamma^5}} = 1.092 \pm 0.032 \nonumber \\
&\!\!\!\!\!M^0_{\gamma^4\phantom{\gamma^5}} = 0.977 \pm 0.015 \ \   
&\phantom{\!\!\!M^0_{\gamma^4\phantom{\gamma^5}}} \nonumber 
\end{eqnarray}
and their first derivatives
\begin{eqnarray}
&&\!\!\!\!\!\!\!\!\!\!\!M'_{\gamma^0\gamma^5} = 
(-0.022, -0.007, -0.008, 0.095, -0.036) \nonumber \\
&&\!\!\!\!\!\!\!\!\!\!\!M'_{\gamma^3\gamma^5} = 
(-0.040, -0.062, -0.020, 0.108, -0.007) \nonumber \\
&&\!\!\!\!\!\!\!\!\!\!\!M'_{\gamma^1\gamma^2} = 
(-0.005, -0.026, -0.016, 0.078, -0.035) \nonumber \\
&&\!\!\!\!\!\!\!\!\!\!\!M'_{\gamma^3\gamma^4} = 
(-0.036, -0.074, -0.029, 0.103, -0.010) \nonumber \\
&&\!\!\!\!\!\!\!\!\!\!\!M'_{\gamma^3} =
(+0.015, +0.006, -0.014, 0.071, -0.019) \nonumber \\
&&\!\!\!\!\!\!\!\!\!\!\!M'_{\gamma^4} = 
(-0.022, -0.074, -0.034, 0.067, -0.022) \nonumber 
\end{eqnarray}

One should notice that the signs of the first derivatives are consistent
for the different pions, but they are very small. They are of the same 
order of magnitude as the error on $M^0_\xi$. This means that within a more 
than reasonable range ($c_i \in [-1,+1]$) the pion's splitting does not vary more than $2\sigma$ where $\sigma$ is its statistical error.

We conclude that the flavor nondegeneracy of the tadpole improved
tree level improved staggered action may be somewhat improved by a 
nonperturbative tuning of the coefficients,  by perhaps $10 \%$ or less.
However, for further dramatic reduction in flavor breaking, new terms
must be added to the action, as in Ref.~\cite{tro01}.

\vskip 0.5cm
This work was performed at Fermilab (U.S. Department of Energy Lab 
(operated by the University Research Association, Inc.), under 
contract DE-AC02-76CHO3000.

\end{document}